\documentclass[12pt]{spieman}  % 12pt font required by SPIE;
\usepackage{amsmath,amsfonts,amssymb}
\usepackage{graphicx}
\usepackage{setspace}
\usepackage{tocloft}
\usepackage{lineno}
\usepackage[numbers]{natbib}
\usepackage{aas_macros}
\title{Measuring metal-sulfides in interstellar dust with PRIMA}

\author[a,*]{Izaskun Jim\'enez-Serra}
\author[b]{Shaoshan Zeng}
\author[b]{Yao-Lun Yang}
\author[a]{Ang\`ele Taillard}
\author[a]{Marta Rey-Montejo}
\author[a]{Laura Colzi}
\author[b]{Nami Sakai}
\author[a]{Asunci\'on Fuente}
\affil[a]{Centro de Astrobiolog\'{\i}a (CAB), CSIC-INTA, Ctra. de Ajalvir km 4, ES--28850, Torrej\'on de Ardoz, Spain}
\affil[b]{Star and Planet Formation Laboratory, Cluster for Pioneering Research, RIKEN, 2-1 Hirosawa, Wako, Saitama, 351-0198, Japan}

\cftpagenumbersoff{figure}
\cftpagenumbersoff{table} 
\begin{document} 
\maketitle

\begin{abstract}
{Sulfur is known to undergone severe depletion} when moving from diffuse clouds to the denser regimes of the interstellar medium in molecular clouds. The form in which sulfur gets depleted onto dust grains, however, remains a mystery. One possibility is that sulfur gets locked in interstellar dust in the form of sulfide minerals. Recently, metal-sulfides such as NaS and MgS have been detected in a shocked molecular cloud in the Galactic Center, suggesting that these molecules could represent an important reservoir of sulfur in dust grains. In this contribution, we discuss the prospect of observing metal-sulfides such as MgS and FeS in absorption experiments carried out with the FIRESS instrument onboard PRIMA using its low-resolution observing mode. Our estimates show that the molecular bands of MgS and FeS found between 20 and 50$\mu m$ could be detected in absorption with S/N$\geq$5  {for sources brighter than 200 mJy} in just 1 hour of observing time against low-mass protostellar objects. This science case, therefore, has the potential to unveil the main reservoir of sulfur in interstellar dust, constraining {in what form sulfur is incorporated} into minor bodies of our Solar-system. 

\end{abstract}

% Include a list of up to six keywords after the abstract
\keywords{interstellar medium, dust, sulfur, metal-sulfides, PRIMA}

% Include email contact information for corresponding author
{\noindent \footnotesize\textbf{*}Izaskun Jim\'enez-Serra,  \linkable{ijimenez@cab.inta-csic.es} }

\begin{spacing}{2}   % use double spacing for rest of manuscript

\section{Introduction}
\label{sect:intro}  % \label{} allows reference to this section

Sulfur is one of the elements commonly present in molecules of biological interest together with carbon, hydrogen, oxygen, nitrogen and phosphorus. In the interstellar medium (ISM), sulfur is known to remain mainly in its {ionized atomic form (S$^+$)} in the gas phase in diffuse clouds and in heavily irradiated photon-dominated regions \citep[or PDRs; see][]{Neufeld_2015,Goicoechea_2021} with an abundance close to the solar abundance \citep[i.e. S/H$\sim$1.5$\times$10$^{-5}$;][]{Asplund_2009}.  
When moving to denser regimes of the ISM, however, sulfur undergoes severe depletion. Indeed, the abundances of sulfur-bearing species such as H$_2$S, SO, SO$_2$, or OCS in the gas phase of translucent clouds and star-forming regions \citep{Fuente_2019,Blake_1996,Fuente_2021,Bouscasse_2022,Goicoechea_2006,Anderson_2013,Feng_2020,Taquet_2020}, can only be explained if sulfur is depleted by a factor of $\sim$10, although the depletion factor strongly depends on the environment \citep{Fuente_2023}. In starless cores and protoplanetary disks, sulfur depletion can be even higher, with typical values $\geq$100 \citep{Ruffle_1999,Vastel_2018,Dutrey_2011,Riviere_2020,LeGal2021}. 

It has been hypothesized that the main reservoir of interstellar sulfur could be  {H$_2$S in ices}, since atomic sulfur is expected to easily hydrogenate on the surface of dust grains upon depletion \citep[][]{Millar_1990,Jansen_1995,Ruffle_1999}. However,  {mid-IR observations -- first carried out with the {\it Kuiper Airborne Observatory} (KAO) and the {\it Infared Telescope Facility} (IRTF), and more recently with the {\it James Webb Space Telescope} (JWST) --} have revealed that  {H$_2$S is not detected in interstellar ices with an upper limit $\leq$0.6\% with respect to water \citep[see][]{Mcclure_2023}}; instead, other sulfur-bearing species such as OCS and SO$_2$ have been claimed in the icy mantles of dust grains in dense molecular clouds and star-forming regions \citep[][]{Geballe_1985,Palumbo_1995,Boogert_1997,Mcclure_2023,Rocha_2024}. The estimated abundances of OCS and SO$_2$ ice represent a very small fraction of the total cosmic sulfur abundance ($\sim$0.2\% and $\sim$0.8-4.0\%, respectively), implying that a large fraction of atomic sulfur is incorporated into dust grains in other (so far unknown) forms. This is known as the {\it sulfur depletion problem} \citep[][]{Agundez_2013}.

%In cometary ices H2S is the most abundant S-bearing molecule, with an abundance of up to 1.5\% relative to water (Bockelée-Morvan et al. 2000), or 1.5 × 10-6 relative to H2. The presence of H2S in interstellar ices was inferred from IR observations toward the high-mass protostar W33A (Geballe et al. 1985). However, a robust detection of this molecule in interstellar ices has not been reported yet, probably because of the overlapping of the 3.92 μm IR band of H2S with a strong IR feature of methanol. The upper limit found for this species toward dense clouds and circumstellar regions (0.6–1.6 × 10-6, and 0.04–0.12 × 10-6 relative to H2, respectively; Smith 1991) accounts for only 10\% of the cosmic S abundance.

In the solar system, sulfur has been found in the form of sulfides and sulfates in primitive meteorites of the CI class \citep{Wasson1988}. The Rosetta mission to comet 67P/Churyumov-Gerasimenko measured samples with a near-solar [S]/[O] ratio, revealing that about 80\% of sulfur was in a refractory form \citep{Calmonte2016}. More recently, the sample return mission from the CI asteroid Ryugu has enabled us to confirm the high sulfur abundances in the samples by their direct analysis \citep{Yoshimura2023}. This indicates that sulfur could efficiently be converted  {from gas-phase/volatile atomic form to sulfur-bearing refractory materials} before the formation of the icy comet and CI asteroid.  {In nearby molecular clouds such as Taurus and Perseus, it has recently been shown that sulfur has to be depleted by a factor of 20 in order to explain the observations of a large set of gas-phase sulfur-bearing molecules obtained across these clouds within the GEMS (Gas phase Elemental abundances in molecular CloudS) large-program \citep[][]{Fuente_2023}.  %Although this depletion factor seems to depend on the environment, 
The GEMS results suggest that negatively charged dust grains enhance sulfur depletion when moving deeper into molecular clouds at visual extinctions A$\rm_V$$\geq$3.5 mag, as inferred from the comparison of the observations with extensive astrochemical modelling \citep{Fuente_2023}. Since the amount of sulfur present in ices is small \citep[$\leq$0.6\% with respect to water; see][]{Mcclure_2023}, it is likely that the majority of sulfur in dense and cold molecular clouds resides in interstellar dust in its refractory core.}

Recently, it has been suggested that sulfur allotropes, such as S$_8$, might be an important form of sulfur in ISM dust \citep{Shingledecker2020}. This is because experimental studies have shown that sulfur allotropes can form from UV irradiation of H$_2$S ices \citep{Cazaux_2022} or electron irradiation of H$_2$S and SO$_2$ ices \citep{Mifsud2022}. However, it is unclear whether these allotropes could explain most of the missing sulfur. In fact, while the presence of S$_8$ has been indicated in the Ryugu sample, its abundance is not sufficient to account for the missing sulfur \citep{Aponte2023}.  {In addition, the detection of solid S$_8$ in the ISM remains challenging. As recently shown by \cite{taillard2025}, this sulfur-bearing compound is unlikely to be detected with the JWST even if all sulfur were in this form.}

Alternatively, sulfide-minerals could be an important reservoir of sulfur on dust grains. FeS nano-inclusions have indeed been found in silicate dust \citep{Kohler_2014} and in asteroids \citep{Matsumoto_2020}. These inclusions 
%which could represent at least one-third of the cosmic sulfur without inducing a significant change in the observed mid-IR silicate absorption bands \citep{Kohler_2014}, 
have been proposed to be present in proto-planetary disks  {as observed with the Infrared Space Observatory (ISO) \citep{keller2002},} and they could account for almost 90\% of the elemental sulfur depleted in these objects \citep{Kama_2019}. In the circumstellar medium, the 30$\mu m$ band observed in carbon stars and some planetary nebulae could be explained by MgS dust, which suggests that sulfur may be incorporated into dust grains in the form of metal-sulfides before being injected into the ISM \citep[][]{Forrest_1981,Kimura_2005}. 

 {Recently, gas-phase metal-sulfide molecules} such as NaS and MgS have been discovered in the ISM toward the Giant Molecular Cloud G+0.693-0.027 located in the Galactic Center \citep{Rey-Montejo_2024}.  {This cloud is one of the richest chemical reservoirs in our Galaxy, with nearly 25 new molecular species recently discovered in the gas phase toward this cloud \citep[see][]{jimenez2025}. G+0.693-0.027 does not show any signposts of massive star formation in the form of H$_2$O masers or dust continuum emission \citep{jimenez2020}. Instead, the rich chemistry observed toward this source is believed to be produced by a large-scale shock induced by a cloud-cloud collision \citep{Zeng_2020}. Therefore,} the presence of NaS and MgS in the gas phase toward G+0.693-0.027 is likely due to the sputtering of the refractory part of dust grains by the shock interaction \citep{Rey-Montejo_2024}. The measured gas-phase abundance of NaS and MgS is very low (of a few 10$^{-13}$), eight orders of magnitude lower than the sulfur cosmic abundance. %Future higher-angular resolution observations obtained with the ALMA interferometer, will measure the size of the emitting regions of NaS and MgS, constraining  the actual abundance of these metal-sulfides toward the G+0.693-0.027 molecular cloud. 
%Preliminary astrochemical modeling indicates that these gas-phase abundances are consistent with the depletion of sulfur by a factor of 3 and of Mg by a factor $\geq$10 (Rey-Montejo et al. in prep.), which suggests that sulfides could represent an important reservoir of sulfur in the ISM. 
 {However, it remains unknown whether these gas-phase metal-sulfides are formed via gas-phase chemical reactions or whether they are present in the gas phase because of the sputtering of refractory grain cores in the shock \citep{Rey-Montejo_2024}.} Therefore, it is essential to observe metal-sulfides in interstellar dust to constrain how much sulfur is actually locked into these refractory species. 

 {Far-IR observations (at wavelengths $\geq$15$\,$$\mu$m) have proven to be very useful for the identification of the mineral composition of dust \citep[][]{kemper2002}. The far-IR spectra of minerals are typically rich in prominent and characteristic features that are attributed to vibrations of the crystal \citep{Nuth1985,Kimura_2005}. Therefore, far-IR spectroscopy is a powerful tool for the study of metal sulfides in interstellar dust. In the following Sections, we show how the PRIMA space mission  \cite{glenn2025} is perfectly suited for the characterization of the mineral composition of interstellar dust, and in particular, of its composition in metal-sulfides.}

%\begin{table}[ht]
%\caption{Fonts sizes and styles.} 
%\label{tab:fonts}
%\begin{center}       
%\begin{tabular}{|l|l|} %% this creates two columns
%% |l|l| to left justify each column entry
%% |c|c| to center each column entry
%% use of \rule[]{}{} below opens up each row
%\hline
%\rule[-1ex]{0pt}{3.5ex}  Document entity & Brief description  \\
%\hline\hline
%\rule[-1ex]{0pt}{3.5ex}  Article title & 16 pt., bold, left justified  \\
%\hline
%\rule[-1ex]{0pt}{3.5ex}  Author names & 12 pt., bold, left justified   \\
%\hline
%\rule[-1ex]{0pt}{3.5ex}  Author affiliations & 10 pt., left justified   \\
%\hline
%\rule[-1ex]{0pt}{3.5ex}  Abstract & 10 pt.  \\
%\hline
%\rule[-1ex]{0pt}{3.5ex}  Keywords & 10 pt.  \\
%\hline
%\rule[-1ex]{0pt}{3.5ex}  Section heading & 12 pt., bold, left justified  \\
%\hline
%\rule[-1ex]{0pt}{3.5ex}  Subsection heading & 12 pt., italic, left justified  \\
%\hline
%\rule[-1ex]{0pt}{3.5ex}  Sub-subsection heading & 11 pt., italic, left justified  \\
%\hline
%\rule[-1ex]{0pt}{3.5ex}  Normal text & 12 pt. \\
%\hline
%\rule[-1ex]{0pt}{3.5ex}  Figure and table captions &  10 pt. \\
%\hline 
%\end{tabular}
%\end{center}
%\end{table} 

\section{Absorption measurements of MgS and FeS interstellar dust with PRIMA}
\label{sec:bands}

Infrared observations  {carried out with the KAO and ISO toward the circumstellar material around evolved carbon stars and planetary nebulae,} have shown strong emission features appearing from 30$\mu m$ and upwards \citep{Forrest_1981}. It has been proposed that the carriers of these emission bands are metal-sulfides.  {Laboratory experiments by \cite{Nuth1985} and \cite{Kimura_2005} have indeed shown that MgS is likely the carrier of the 30$\mu m$ emission band seen in evolved stars.} Theoretical calculations simulating dust condensation processes in circumstellar material predict that  {metal sulfides such as MgS and FeS, can condensate in expanding gas shells with temperatures $\leq$1200 K independently on the pressure conditions \citep{Lattimer1978}}.   {This dust is then incorporated into the interstellar medium in molecular clouds, which form the subsequent generation of stars. FeS has been claimed to be present in both planetary nebulae \citep{Hony2002} and protoplanetary disks \citep{keller2002};} and iron sulfides are known to be an important constituent of minor Solar-system bodies \citep{Kerridge1976,Lawler1989,Dai2001}, which suggests that MgS and FeS are likely present in the refractory part of interstellar dust grains. 

Unfortunately, observations of MgS and FeS in dust in the ISM are lacking. This is due to the fact that observations of these refractory materials in interstellar dust in molecular clouds have to be carried out in absorption against the strong continuum emission of bright background sources. MgS and FeS are known to present several molecular bands between 25 and 80$\mu m$ \citep{Hony2002,Kimura_2005}.  {However, these bands are difficult to observe in absorption since they require high sensitivity observations \citep{Hofmeister2003}. Despite having similar mass absorption coefficients in the IR to metal oxides such as SiO, the IR bands of metal sulfides such as FeS and MgS are indeed expected to be at least 40 times weaker than those of metal oxides as a result of the large difference between the cosmic abundance of elemental oxygen (4.9$\times$10$^{-4}$ with respect to hydrogen) and the abundance of elemental sulfur \citep[of 1.5$\times$10$^{-5}$; see][]{Asplund_2009}.} 

\begin{figure}
\begin{center}
\begin{tabular}{c}
\includegraphics[height=5.5cm]{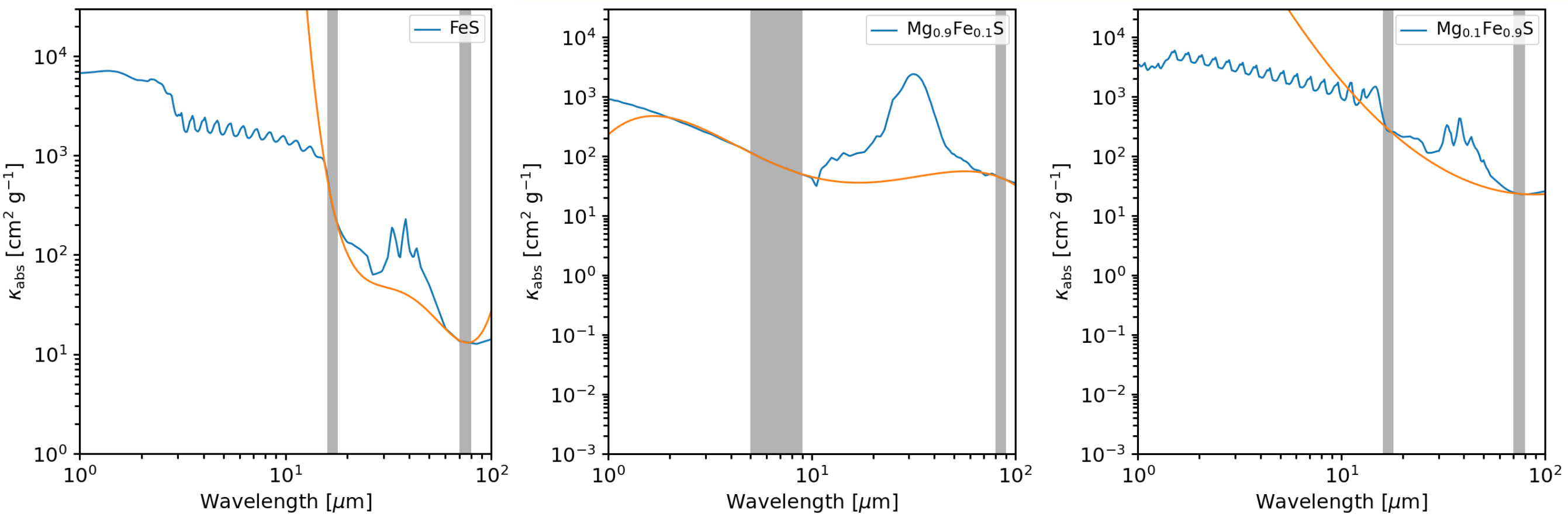}
\end{tabular}
\end{center}
\caption{Synthesized dust opacity profiles from 1 to 100$\mu m$ obtained with the {\it optool} software for  {pure samples of} FeS (left panel) and by using the optical constants of Mg$_x$Fe$_{1-x}$S (with x=0.9 and 0.1 for the middle and right panels, respectively) extracted from the JPDOC database. Vertical grey areas indicate the parts of the spectrum used to perform the local baseline of the observed FeS and Mg$_x$Fe$_{1-x}$S bands between 10 and 100$\mu m$ (see orange lines).}
\label{figure1}
\end{figure}

\begin{figure}
\begin{center}
\begin{tabular}{c}
\includegraphics[height=6.2cm]{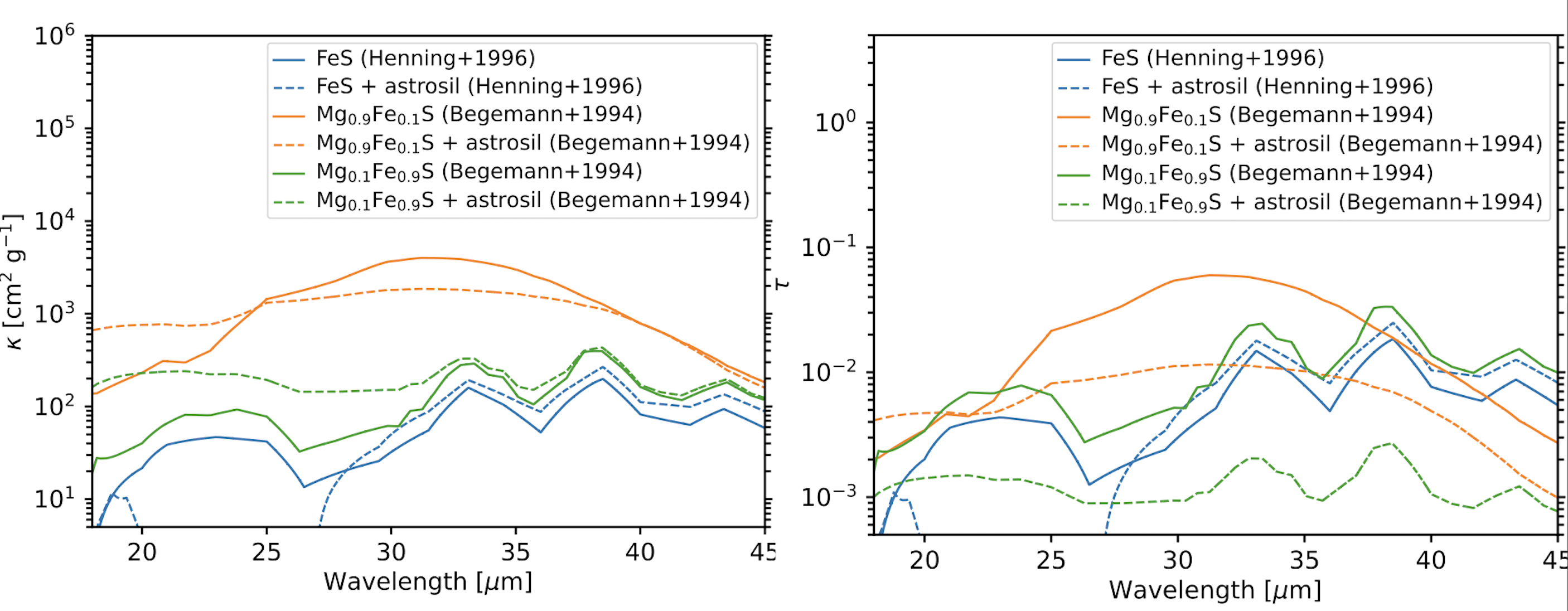}  % fig2 includes two images 
%(a) \hspace{5.1cm} (b)
\end{tabular}
\end{center}
\caption 
{\label{figure2}
{\it Left panel}: Zoom-in of the synthesized dust opacity profiles of 100\% FeS (blue line), of 100\% Mg$_{0.1}$Fe$_{0.9}$S (green line), and of 100\% Mg$_{0.9}$Fe$_{0.1}$S (orange line)  {for pure samples (solid lines) and with astrosilicates mixtures (dashed lines). The profiles with astrosilicates mixtures have been calculated using the dust model of \cite{draine2003}.}  {Right panel}: Optical depth for the FeS and MgS molecular bands expected between 20 and 50$\mu m$  {for the pure samples (solid lines) and the astrosilicates mixtures (dashed lines).} The opacity and optical depth profiles have been smoothed to the expected resolving power of R$>$85 of the low-resolution mode of the instrument FIRESS onboard PRIMA.} 
\end{figure}

In Figure$\,$\ref{figure1}, we show the molecular bands of FeS and MgS as seen in the synthesized dust opacity profiles of FeS and of Mg$_x$Fe$_{1-x}$S obtained with the {\it optool} software \citep[][]{Dominik2021} with a built-in model of FeS \citep[][]{Henning1996}, and by using the JPDOC database (see https://www.astro.uni-jena.de/Laboratory/Database/jpdoc/f-dbase.html.) for the optical constants of Mg$_x$Fe$_{1-x}$S, which uses the data from \cite{Begemann_1994}. In these models, dust is assumed to be  {pure samples of} 100\% FeS (left panel, Figure$\,$\ref{figure1}), 100\% Mg$_{0.9}$Fe$_{0.1}$S (middle panel), and 100\% Mg$_{0.1}$Fe$_{0.9}$S (right panel).  {In these calculations, we have assumed a MRN grain size distribution that follows a power-law profile $\propto a^{-3.5}$, with $a$ being the grain size varying from 0.1 to 1$\mu m$ \citep{mathis1977}. A size of 0.1$\mu m$ is the typical size measured in the ISM \citep{draine2003}, while dust grain sizes of $\sim$1$\mu m$ have recently been measured in dense molecular clouds with the JWST \citep{Mcclure_2023}. %The synthesized dust opacity profiles from 1$\mu m$ to 100$\mu m$ are presented in Figure$\,$\ref{figure1}. 
The FeS and Mg$_{0.1}$Fe$_{0.9}$S profiles show several clear narrow fearures between 30 and 50$\mu m$, while the Mg$_{0.9}$Fe$_{0.1}$S profile presents a broad IR band peaking at $\sim$30$\mu m$.
A local baseline was fit around the FeS, Mg$_{0.9}$Fe$_{0.1}$S, and Mg$_{0.1}$Fe$_{0.9}$S features, by anchoring them using the feature-free wavelength ranges shown by gray shaded areas in Figure$\,$\ref{figure1}. The resulting profiles are reported with solid lines in Figure$\,$\ref{figure2} (left panel).}

In Figure$\,$\ref{figure2}, we present a zoom-in of the synthesized dust opacity profiles  {obtained using the expected resolving power of} the low-resolution mode of  {PRIMA's far-IR enhanced survey spectrometer \citep[FIRESS; see][]{bradford2025}.} FIRESS uses four long-slit grating spectrometer modules to cover the full 24 to 235$\mu$m waveband  {with a resolving power R$>$85. This resolving power corresponds to spectral resolutions between 0.35-0.5$\mu m$ at the wavelengths of the FeS, Mg$_{0.9}$Fe$_{0.1}$S, and Mg$_{0.1}$Fe$_{0.9}$S bands.} From Figure$\,$\ref{figure2}, it is clear that the resolving power of the FIRESS low-resolution mode is sufficient to resolve the series of  {narrow} bands of FeS between 30 and 50$\mu m$, which will allow us to distinguish between FeS-dominated from MgS-dominated interstellar dust. 

 {We note, however, that interstellar dust grains are unlikely to be pure FeS or Mg$_x$Fe$_{1-x}$S. Instead, these metal-sulfides are likely mixed with silicates or carbonaceous material, which may significantly alter the observed dust opacity profiles. To illustrate this effect, in Figure$\,$\ref{figure2} (dashed lines) we also report the dust opacity profiles of FeS, Mg$_{0.9}$Fe$_{0.1}$S, and Mg$_{0.1}$Fe$_{0.9}$S synthesized by mixing them with the astrosilicate dust model from \citep{draine2003}. The assumed mixing ratio is 1:1 by mass, but note that this is an extreme case since the amount of metal sulfides is likely a fraction of the amount of silicates in interstellar dust \citep[$\leq$60\%; see][]{keller2002}. However, we opt for making the calculations for this large mixing ratio to show the maximum impact on the broadening of the metal-sulfide IR bands. As seen from Figure$\,$\ref{figure2} (left panel), the FeS IR bands between 30 and 50$\mu m$ obtained with the astrosilicates mixtures remain narrow and they can be resolved using the low-resolution mode of the FIRESS instrument (i.e. there is little difference between the pure samples and the mixtures). The FeS IR band at 23.5$\mu m$ and the MgS IR band at 30$\mu m$, however, experience some broadening in the astrosilicates mixtures but, as demonstrated in Figure$\,$\ref{figure2b} and Section$\,$\ref{sec:time} below, this is not expected to prevent the detection of the MgS IR band at 30$\mu m$.}

In order to calculate the time estimates, we need to convert dust opacities into optical depths. To do so, we have to multiply the synthesized dust opacity profiles shown in Figure$\,$\ref{figure2} (left panel) by the mass column density of FeS and MgS. We assume that 50\% of sulfur is locked into FeS, yielding an abundance of 6.6$\times$10$^{-6}$ for FeS.  {Several estimates of the amount of the cosmic sulfur abundance locked into FeS refractory materials are available from the literature and they range between 30\% \citep[from the study of infrared extinction curves;][]{koehler2014} to 90\% \citep[from a sample of young stars with accretion-contaminated photospheres;][]{Kama_2019}. We assume an intermediate value of 50\% for FeS, which is consistent with the value obtained by \cite{Kama_2019} for their sample of young stars when using the solar abundances from \cite{Asplund_2009} as the stellar reference. 

For MgS, we assume that only 5\% of sulfur is locked into refractory MgS, which corresponds to an abundance of 6.6$\times$10$^{-7}$. This value is derived from the gas-phase abundance of MgS estimated toward the Galactic Center molecular cloud G+0.693 assuming a source size of 2$"$ \citep{Rey-Montejo_2024} and a typical depletion factor of 100. Note that an abundance of 6.6$\times$10$^{-7}$ for refractory MgS} should be considered as a lower limit since the Mg/Fe ratio in the solar system is $\sim$0.5. Therefore, if a larger fraction of S were incorporated into MgS, its molecular band at $\sim$30$\mu m$ would be detected with a higher S/N than estimated here. 
From all this, and assuming a column density of hydrogen nuclei of N(H)=10$^{23}$ cm$^{-2}$, the derived mass column density for FeS is 9.3$\times$10$^{-5}$ g$\,$cm$^{-2}$ and 6.2$\times$10$^{-6}$ g$\,$cm$^{-2}$ for MgS. By using these values, we obtain the optical depths shown in Figure$\,$\ref{figure2}, right panel. 

\section{Time estimates with PRIMA}
\label{sec:time}

\begin{figure}
\begin{center}
\begin{tabular}{c}
\includegraphics[height=7cm]{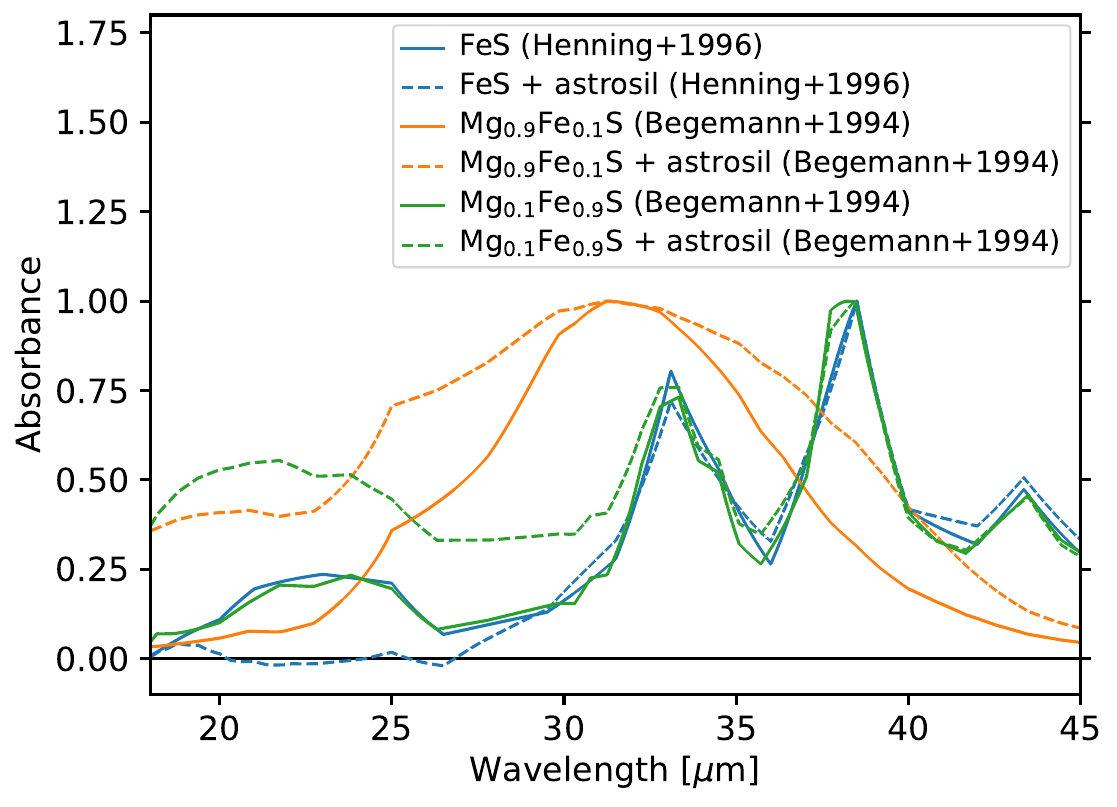}
\end{tabular}
\end{center}
\caption 
{\label{figure2b}
 {Zoom-in of the absorbance profiles of 100\% FeS (blue line), 100\% Mg$_{0.1}$Fe$_{0.9}$S (green line), and 100\% Mg$_{0.9}$Fe$_{0.1}$S (orange line) for pure samples (solid lines) and with astrosilicates mixtures (dashed lines). The astrosilicates mixtures profiles have been calculated using the dust model of \cite{draine2003}. The absorbance profile has been synthesized using the expected resolving power of R$>$85 of the low-resolution mode of the instrument FIRESS onboard PRIMA.}} 
\end{figure}

 {To study the feasibility of detection of the IR molecular bands of FeS and MgS from interstellar dust with PRIMA, we design an experiment in which these bands are observed in absorption against the bright continuum emission of a background source. To estimate the intensity of the observed absorption bands, we use as input the optical depths shown in Figure$\,$\ref{figure2} (right panel). The absorbance profiles of the pure samples of FeS, Mg$_{0.9}$Fe$_{0.1}$S, and Mg$_{0.1}$Fe$_{0.9}$S, and of their mixtures with astrosilicates, are shown in Figure$\,$\ref{figure2b}. The dust optical depth ($\tau$) and the absorbance ($A$) are related as $\tau \; $=$\; \ln{(\mathcal{F}_0 / \mathcal{F}}) \; = \; A \ln(10)$, with $\mathcal{F}$ the observed flux and $\mathcal{F}_0$ the original flux of the background source.  

From Figure$\,$\ref{figure2b}, one can find that the dominant grain composition in either FeS or MgS will clearly be probed by FIRESS, given the obvious differences between the spectral profiles of FeS:astrosilicates grains and MgS:astrosilicates dust. In particular, the narrow IR bands of FeS between 30 and 50$\mu m$ do not show significant differences with respect to the FeS:astrosilicates mixtures, which indicates that their detection will represent a robust signature of the presence of FeS in interstellar dust. In addition, note that the IR bands probed by PRIMA are expected to be factors $\sim$2-4 more intense than the 23.5$\mu m$ band and, although this band is currently accessible with JWST, JWST's sensitivity significantly worsens from 20$\mu$m longwards by factors 4-100, which hampers its detection. 
For the MgS mixtures, despite the broadening of the 30$\mu m$ band, this band will be observed in full by the wavelength coverage of the FIRESS instrument (24-235$\mu m$) onboard PRIMA.} 

\begin{figure}
\begin{center}
\begin{tabular}{c}
\includegraphics[height=9cm]{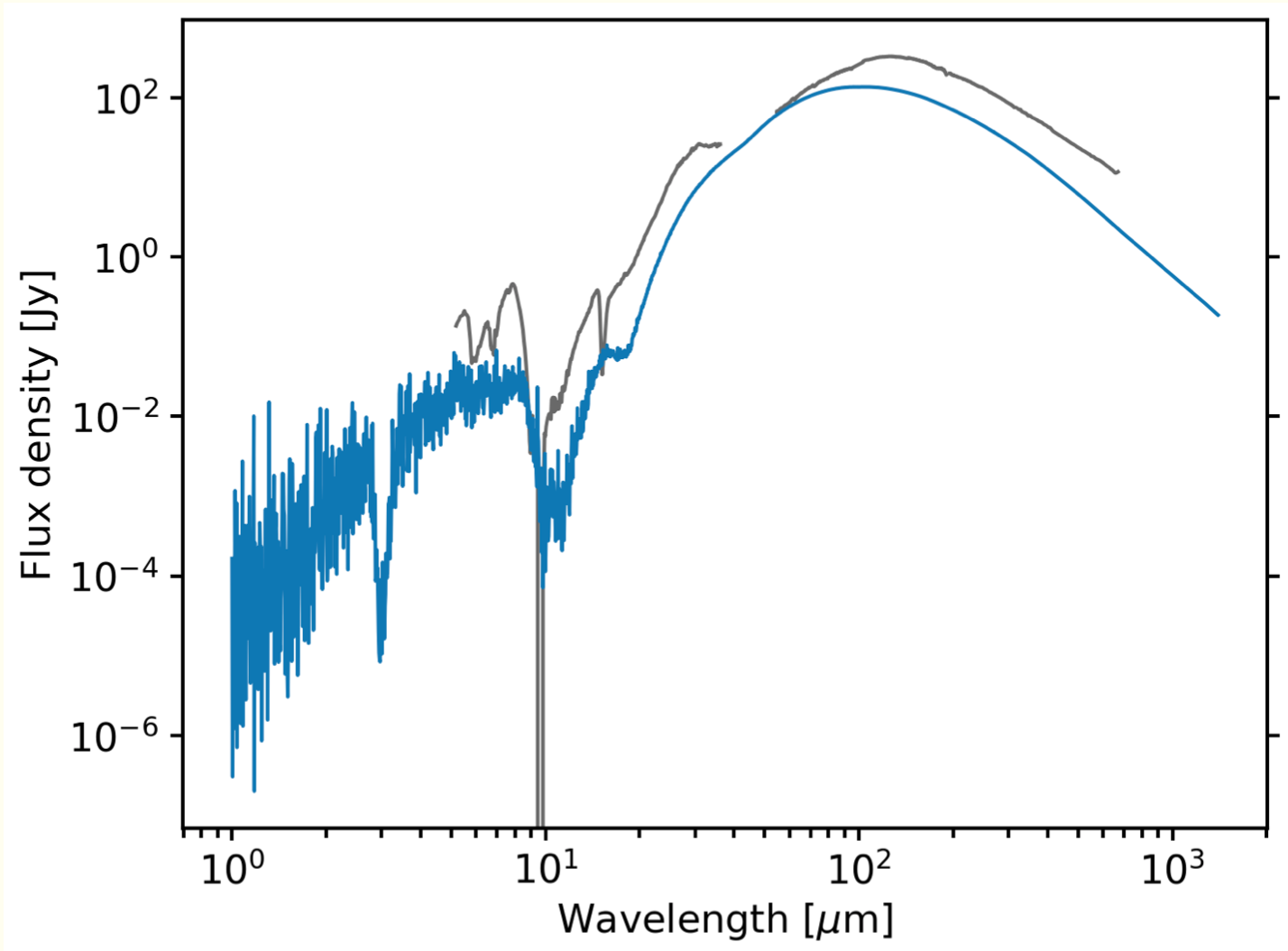}  
\end{tabular}
\end{center}
\caption 
{\label{figure4}
Synthetic SED modelled by \cite{Yang_2017} for the BHR 71 YSO. BHR 71 is a low-mass protostar with a luminosity of $\sim$10 L$_\odot$. The gray data are from Spitzer IRS, Herschel PACS and SPIRE. The synthetic SED (blue) is extracted from a 7.6$"$ aperture,  {which is the pixel size of the FIRESS instrument at 30$\mu m$}.} 
\end{figure}

 {With this information, we can now estimate the time required to detect the IR absorption bands of FeS and MgS with S/N$\geq$5. As an example, we can use the BHR 71 Young Stellar Object (YSO) source, a typical $\sim$10 L$_\odot$ low-mass protostar, which has been modelled by \cite{Yang_2017} (see Figure$\,$\ref{figure4}). For the MgS:astrosilicates mixtures,} the optical depth of Mg$_{0.9}$Fe$_{0.1}$S at $\sim$30$\mu m$ (i.e. at the peak of the band) is $\tau$=0.01 (see right panel of Figure$\,$\ref{figure2}).  {From Figure$\,$\ref{figure4},} the observed continuum flux of BHR71 at $\sim$30$\mu m$ is $\sim$3 Jy. The intensity of the MgS absorption band,  {I$_{\rm abs}$, can then be calculated as I$_{\rm abs}$ = I$_{\rm cont}$$\times$(1 - e$^{\tau({\rm MgS})}$), with I$_{\rm cont}$ the flux of the background source and $\tau_{\rm MgS}$ the optical depth of the Mg$_{0.9}$Fe$_{0.1}$S band at $\sim$30$\mu m$. Assuming I$_{\rm cont}$=3 Jy and $\tau_{\rm MgS}$=-0.01, we derive an intensity for the absorption feature of I$_{\rm abs}$ = 30 mJy.}

The 5$\sigma$ sensitivity of the FIRESS instrument for a point source in its low-resolution mode (R$>$85) in 1 hour of integration time is 1.9$\times$10$^{-19}$ W/m$^2$,  {as indicated in the sensitivity requirements for FIRESS (see https://prima.ipac.caltech.edu/page/instruments). This is equivalent to 0.16 mJy.} For the unit conversion, we have used the expression Flux(mJy) = [Flux (W/m$^2$) $\times$ R $\times$ $\lambda$($\mu m$)]/3$\times$10$^{-15}$, with R the resolving power of the instrument, and $\lambda$ the wavelength of the observed IR band  {(see Appendix A in the Spitzer Telescope Handbook; https://irsa.ipac.caltech.edu/data/SPITZER/docs/spitzermission/missionoverview/spitzertelescopehandbook/18/).}
%(see slide 3 in https://prima.ipac.caltech.edu/page/instruments)
This implies that the MgS IR absorption band at 30$\mu m$ will be detected with a S/N$\sim$940. We note that the high-resolution observing mode of FIRESS (R=1180 at 30$\mu m$), would require 74 hours of integration time  {for a 5$\sigma$ detection}. However, the resolving power of the FIRESS low-resolution mode (R$>$85), is sufficient to resolve the broad MgS feature seen between 20 and 50$\mu m$ (see Figures$\,$\ref{figure2} and \ref{figure2b}). 

%I have extracted the value of R for FIRESS low-res mode from here: https://prima.ipac.caltech.edu/download/MediaFile/88/binary/original

%For the high-res mode (R=4400 x (lambda/112 micron) = 1180 at 30 micron), the 5 sigma sensitivity would be 236 mJy, which would require about 74 hours of integration time. But I think we can definitely go ahead with the low-res mode, for which we will be able to resolve the band.

 {For the FeS:astrosilicates mixtures, the optical depth of the weakest IR band feature between 20 and 50$\mu m$ is $\sim$10$^{-3}$  (see Figure$\,$\ref{figure2}, right panel).} The expected intensity of the FeS absorption bands will therefore be  {I$_{abs}$ = 3 Jy$\times$(1 - e$^{-0.001}$) = 3 mJy.}
The 5$\sigma$ sensitivity of the FIRESS instrument for a point source in its low-resolution mode (R=85) is again 0.16 mJy, which implies that the FeS IR absorption bands between 20 and 50$\mu m$ can be detected  {with S/N$\sim$90} in just 1 hour of integration time.  {We note that these time estimates do not depend on the telescope aperture since the absorption spectra is obtained against the background YSO, which for the pixel-size of FIRESS of 7.6$"$ is a point source.}

\begin{figure}
\begin{center}
\begin{tabular}{c}
\includegraphics[height=11cm]{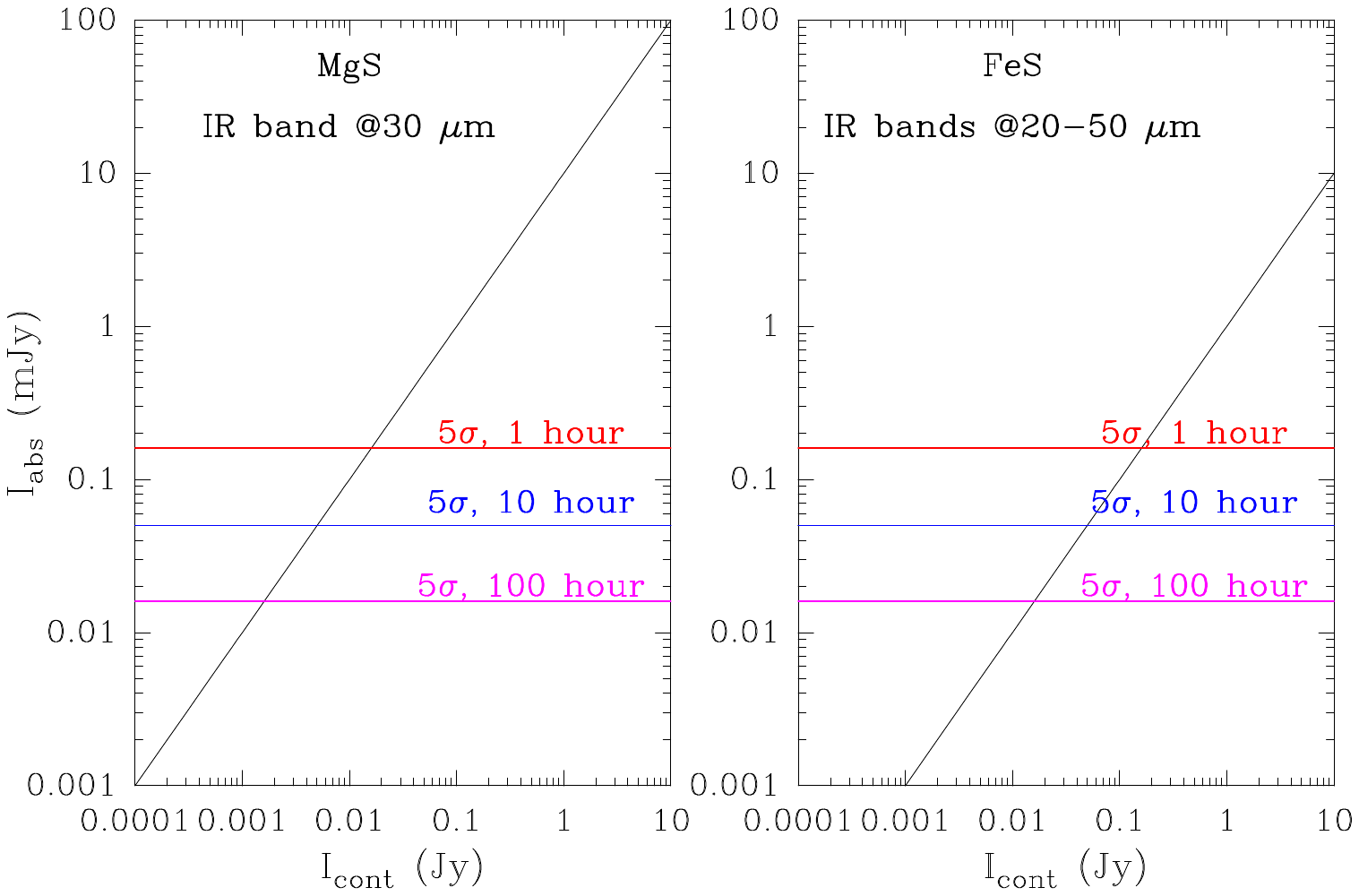}  
\end{tabular}
\end{center}
\caption 
{\label{figure5}
 {Predicted intensity of the IR absorption bands of FeS and MgS mixtures with astrosilicates (I$_{abs}$) observed against a background source of brightness I$_{cont}$ (see black lines). Red, blue and magenta horizontal lines indicate the 5$\sigma$ detection thresholds for 1 hour, 10 hours and 100 hours of integration time, respectively.}} 
\end{figure}

With this modest investment of time per source, we foresee observations of a collection of  {targets ranging from dense molecular clouds to Class 0, Class I and Class II YSO sources located in different environments} \citep[with different values of the external radiation field, as explored by][]{Fuente_2023}, to understand whether metal-sulfides represent the main reservoir of sulfur in interstellar dust, and in what amount under the different physical environments. 
 {The feasibility of these studies is reported in Figure$\,$\ref{figure5}, which presents the predicted intensity of the IR bands of FeS and MgS mixed with astrosilicates as a function of the background source flux. From Figure$\,$\ref{figure5}, we find that the MgS IR band at 30$\mu m$ will be detected with a S/N$\geq$5 toward sources brighter than 20 mJy in 1 hour of integration time; while for the FeS IR bands, a 5$\sigma$ detection will require sources brighter than 200 mJy for the same integration time.}  
This will allow the mapping of metal-sulfides in their refractory form across different extinction regimes in molecular clouds. 

In addition to metal-sulfides, we have also explored the possibility of targeting allotropes such as S$_8$ with PRIMA, as alternative reservoirs of sulfur in the ISM \citep{Cazaux_2022}. S$_8$ indeed presents three IR bands at 21.1$\mu m$, 41.3$\mu m$, and 53.4$\mu m$ \citep{Ferrari_2024}.  {The S$_8$ band at 21.1$\mu m$, however, cannot be detected with the JWST as its band strength is too weak and the band is too wide \citep{taillard2025}.} Our estimates of the intensity of the IR bands at 41.3$\mu m$ and 53.4$\mu m$ gives  {$\sim$0.01 mJy (or $\sim$8.8$\times$10$^{-21}$ W/m$^2$), which would require over 500 hours of integration time with the FIRESS instrument onboard PRIMA in its low-resolution mode for a 5$\sigma$ detection, clearly lying outside the detectability range of PRIMA.} Therefore, metal-sulfides represent an excellent target for this future space mission.

 {Finally, we would like to stress that these observations will represent a huge improvement with previous observations of these IR bands obtained with ISO. The inspection of the far-IR spectra measured toward a sample of proto-planetary disks by \cite{keller2002} and \cite{vandenancker2000}, shows that the rms noise level is at best 2 Jy for an on-target integration of 3454 sec (nearly 1 hour) after heavily smoothing the spectra to a similar spectral resolution (R=100) to that provided by FIRESS in its low-resolution mode. This implies that the PRIMA observations will be $\geq$10$^4$ times more sensitive than those obtained with ISO for similar integration times (note that the 5$\sigma$ sensitivity of the FIRESS instrument in its low-resolution mode for 1 hour integration is 0.16 mJy).}

\section{Conclusions}

In this contribution, we have explored the detectability of metal-sulfides such as FeS and MgS in interstellar dust. These sulfides, especially MgS, which has been tentatively detected toward evolved stars, could represent an important reservoir of sulfur in the ISM. 
 {Our contribution shows that PRIMA will be a key instrument in the detection of metal-sulfides in interstellar dust for the following reasons: 

\begin{itemize}
\item i) the IR band of MgS at 30$\mu$m cannot be accessed with the JWST since JWST's wavelength coverage goes up to 28$\mu $m; 

\item ii) FeS presents several narrow IR bands between 30$\mu$m and 50$\mu$m, which not only are expected to be brighter than its IR band at 23.5$\mu$m (by factors $\sim$2-4), but they should be easier to recognize in astronomical data \citep[see][and Figure$\,$\ref{figure2b}]{Hofmeister2003}.
\end{itemize}}

Using reasonable assumptions for the abundance of FeS and MgS in interstellar dust  {(including mixtures with astrosilicates)}, our time estimates indicate that the absorption band of MgS at 30$\mu m$ could be detected with a S/N$\sim$940 in just 1 hour of integration time with the FIRESS instrument onboard PRIMA in its low-resolution mode, while the FeS bands between 20 and 50$\mu m$ could be detected with S/N$\geq$90 for the same integration time with the same instrument and spectral resolution mode. These observations will be essential to establish the abundance of metal-sulfides in interstellar dust and, hence, to constrain how much of the elemental abundance of sulfur is locked into sulfide minerals, potentially unveiling the main reservoir of sulfur in the ISM and helping to solve the sulfur depletion problem. 

\section{Disclosures}

The authors declare that there are no financial interests, commercial affiliations, or other potential conflicts of interest that could have influenced the objectivity of this research or the writing of this paper. 

\section{Code, Data, and Materials Availability}

The dust opacity profiles of FeS shown in Figures$\,$\ref{figure1} and \ref{figure2}, are available on GitHub within the {\it optool} software \citep{Dominik2021} at https://github.com/cdominik/optool. The dust opacity profiles of Mg$_x$Fe$_{1-x}$S are available within the JPDOC database (see https://www.astro.uni-jena.de/Laboratory/Database/jpdoc/f-dbase.html).

\section{Acknowledgments}
 I.J.-S. acknowledges funding from the ERC grant OPENS (project number 101125858) funded by the European Union. Views and opinions expressed are however those of the author(s) only and do not necessarily reflect those of the European Union or the European Research Council Executive Agency. Neither the European Union nor the granting authority can be held responsible for them. I.J-.S also acknowledges funding from grant PID2022-136814NB-I00 funded by the Spanish Ministry of Science, Innovation and Universities/State Agency of Research MICIU/AEI/ 10.13039/501100011033 and by “ERDF/EU”. I.J.-S, S. Z., L.C., and N.S.  also acknowledge support from the CSIC ILINK project SENTINEL (ILINK23017). S. Z acknowledges the support by RIKEN Special Postdoctoral Researchers Program. S.Z, Y.-L.Y., and N. S also acknowledge funding by a Grant-in-Aid from Japan Society for the Promotion of Science (KAKENHI: Nos. 20H5844 and 20H5845). A.F. and A.T. acknowledges funding from the European Union (ERC, SUL4LIFE, grant agreement No101096293). A.F. also thanks project PID2022-137980NB-I00 funded by the Spanish Ministry of Science and Innovation/State Agency of Research MCIN/AEI/10.13039/501100011033 and by "ERDF A way of making Europe".

%\subsection{Biographies}
%A brief professional biography of approximately 75 words may be provided for each author, if available. Biographies should be placed at the end of the paper, after the references. Personal information such as hobbies or birthplace/birthdate should not be included. Author photographs are not published.

%\section{Figures and Tables}

%\appendix    % this command starts appendixes

%\section{Miscellaneous Formatting Details}
%\label{sect:misc}
%At times it may be desired, for formatting reasons, to break a line without starting a new paragraph. In a LaTeX source file, a linebreak is created with \verb|\\|.

%%%%% References %%%%%

\bibliography{report}   % bibliography data in report.bib
\bibliographystyle{spiejour}   % makes bibtex use spiejour.bst

%%%%% Biographies of authors %%%%%

\vspace{2ex}\noindent\textbf{Izaskun Jim\'enez-Serra} is a chief scientist at the Center of Astrobiology (CAB) in Madrid (Spain). She received her BS, MS, and PhD degrees in physics and astrophysics from the Spanish Universidad Complutense de Madrid in 2002, 2003, and 2007 respectively. She worked at different international research centers and institutions as Submillimeter Array fellow (USA), Marie-Curie fellow (Germany) and STFC Ernest Rutherford fellow (UK). She is the author of more than 210 journal papers and has written four book chapters. Her research interests include astrochemistry, star-formation and the search of prebiotic species in the ISM. She is the PI of the ERC Consolidator Grant OPENS (the Onset of Prebiotic chEmistry iN Space; project number 101125858).

\vspace{2ex}\noindent\textbf{Shaoshan Zeng} is currently a postdoctoral researcher at RIKEN in Japan. She obtained her BS and MS degrees in University College London (UK) and her PhD degree in Queen Mary University of London (UK). Her main research interest is astrochemistry, focus mainly on complex organic molecules in the ISM, including the Galactic Centre molecular cloud and star-forming regions in the Galactic disc. She has also been studying the kinematics and origin of shock towards the G+0.693-0.027 molecular cloud.

\vspace{2ex}\noindent\textbf{Yao-Lun Yang} is a research scientist at the Star and Planet Formation Laboratory in Cluster for Pioneering Research in RIKEN.  He received his BS in physics from National Taiwan University in 2012 and his MS and PhD in astronomy from the University of Texas at Austin in 2015 and 2019, respectively.  He worked as JSPS postdoctoral fellow in RIKEN and Origin postdoctoral fellow at the University of Virginia prior to his current position.  His research focuses on characterizing physical and chemical processes in star-forming regions, using multi-wavelength observations.  He is also the lead of Star and Planetary Science in the JAXA/ISAS Research Group for the FIR Probe mission.

\vspace{2ex}\noindent\textbf{Angèle Taillard} is a postdoctoral researcher at the Center of 
Astrobiology (CAB). She obtained her BS, MS and PhD degrees by the 
University of Bordeaux, France. She is an expert in chemical modeling 
of star forming regions and has developed a synthetic ice spectra 
generator to study new possible detections in infrared.

\vspace{2ex}\noindent\textbf{Marta Rey-Montejo} is a master student at the Spanish Universidad Completense de Madrid and at the Center of Astrobiology (CAB), supervised by I. Jim\'enez-Serra. She has recently reported the detection of the metal-sulfides NaS and MgS in the gas phase toward the Galactic Center molecular cloud G+0.693-0.027. She is currently carrying out the astrochemical modelling of these metal-sulfides to constrain how much of the elemental abundance of sulfur is depleted into interstellar dust.  

\vspace{2ex}\noindent\textbf{Laura Colzi} is a postdoctoral researcher at the Center of Astrobiology (CAB). She obtained her BS, MS and PhD degrees by the University of Florence. She is currently leading projects to study isotopic ratios at different spatial scales across galactic and extragalactic environments, both from an observational and a theoretical point of view. She is also an expert in the physical properties of the molecular clouds in the Galactic Center, such as G+0.693-0.027. 

\vspace{2ex}\noindent\textbf{Nami Sakai} is a chief scientist at RIKEN in Saitama, Japan, and the director of the Star and Planet Formation Laboratory within the Cluster for Pioneering Research at RIKEN.  She reveived her BS degree in physics from Waseda University in 2004 and her MS and PhD degrees in physics from the University of Tokyo in 2006 and 2008 respectively.  After working as an assistant professor at the same university, she assumed her current position.  She is the author of more than 160 refereed journal papers and has written 3 book chapters.  Her research interests are star and planet formation, chemical evolution and diversity of star and planet forming regions, and development of microwave spectrometer for molecules in astronomical interests. She is the PI of the Grant-in-Aid for transformative Research Areas (A) Next Generation Astrochemistry, supported by MEXT/JSPS(KAKENHI) (project number 20H05844).

\vspace{2ex}\noindent\textbf{Asunci\'on Fuente} is currently working at the CAB with a distinguished researcher contract. Her main research area is astrochemistry, and she has been heavily involved in the design of the observational programmes of the several millimetre and infrared telescopes (Herschel, ALMA, JWST). She is the principal investigator of the project ‘The trail of sulphur: from molecular clouds to life’ (SUL4LIFE), funded by the European Unnion through an ERC-AdG-2022 (Grant Agreement No. 101096293). The aim of this project is to determine precisely the evolution of the sulphur abundance in the interstellar medium until its incorporation into life.

%\vspace{1ex}
%\noindent Biographies and photographs of the other authors are not available.

\listoffigures
\listoftables

\end{spacing}
\end{document}